\documentclass[useAMS,usegraphicx]{mn2e}
\usepackage[]{graphicx}
\usepackage{longtable}
\usepackage{setspace}

\newcommand{\xte}{\textit{RXTE}}
\newcommand{\uhuru}{\textit{UHURU}}
\newcommand{\ginga}{\textit{GINGA}}

\begin{document}

\title[Cen X-3: Superorbital flux Variations]{Long term flux variations in Cen X-3: clues from flux dependent orbital modulation and pulsed 
fraction}

\author[Raichur \& Paul]{Harsha Raichur$^{1,2}$, Biswajit Paul$^{1}$\\
1. Raman Research Institute, Sadashivanagar, C. V. Raman Avenue,
Bangalore 560\,080, India \\
2. Joint Astronomy Programme, Indian Institute of Science,
Bangalore 560\,080, India\\
(E-mail:sharsha{@}rri.res.in,bpaul{@}rri.res.in)}
\date{Accepted.....; Received .....}
\maketitle

\begin{abstract}

We have investigated the long term flux variation in Cen X-3 using
orbital modulation and pulsed fraction in different flux states using
observations made with the All Sky Monitor 
and the Proportional Counter Array 
on board the {\it Rossi X-ray Timing Explorer}. 
In the high state, the eclipse ingress and egress are found to be
sharp whereas in the intermediate state the transitions are more gradual.
In the low state, instead of eclipse ingress and egress, the lightcurve
shows a smooth flux variation with orbital phase.
The orbital modulation of the X-ray lightcurve in the low state shows 
that the X-ray emission observed in this state is from an extended object.
The flux dependent orbital modulations
indicate that the different flux states of Cen X-3 are 
primarily due to varying degree of obscuration.
Measurement of the pulsed fraction in different flux states is consistent
with the X-ray emission of Cen X-3 having one highly varying 
component with a constant pulsed fraction and an unpulsed component
and in the low state, the unpulsed component becomes dominant.
The observed X-ray emission in 
the low state is likely to be due to scattering of X-rays from the stellar 
wind of the companion star.
Though we can not ascertain the origin and nature of the obscuring
material that causes the aperiodic long term flux variation, we
point out that a precessing accretion disk driven by radiative forces is a
distinct possibility.

\end{abstract}


\section{Introduction}

Several persistent X-ray binaries show large flux variations in their
X-ray lightcurves on timescales significantly longer than their orbital
periods (Wen et al. 2006).
Highly periodic flux variaitons at time scales larger than their
respective orbital periods are seen in
Her X-1 (35 d: Tananbaum et al. 1972, Still \& Boyd 2004),
LMC X-4 (30.28 d: Paul \& Kitamoto 2002),
2S 0114+650 (30.75 d: Farrell, Sood \& O'Neill 2006),
SS 433 (164 d: Eikenberry et al. 2001; Fabrika 2004), 
XTE J1716--389 (99.1 d: Cornelisse, Charles \& Robertson 2006),
4U 1820--303 (172 d: Smale \& Lochner 1992; Zdziarski et al. 2007a)
and Cyg X-1 (150 d: Kitamoto et al. 2000; Lachowicz et al. 2006). 
The first three of these seven sources, known with stable superorbital period
are accretion powered pulsars.
Several other X-ray binaries show quasi periodic long term flux 
variations.
SMC X-1 has a long term period in the range of 50-70 d
(Grubers \& Rothschild 1984; Clarkson et al. 2003, Wen et al. 2006).
GRS 1747-312
shows quasi periodic flux variations of $\sim$ 150 d
\cite{Zand03,Wen06}. 
Cyg X-2 shows quasi periodic long term variability with
period in the range of 60-90 d \cite{Smale92,Paul00} and
LMC X-3 also shows long term flux variations with periodicity in the
range of 100-500 d \cite{Wen06} that is unstable over a longer period
\cite{Paul00,Wen06}.
Long term quasi periodic variations with a period of about 217
d are also seen from the Rapid Burster (X1730--333) which is in the
form of recurrent outburst rather than a gradual change in X-ray
flux \cite{Masetti02}.

Cen X-3 is a high mass X-ray binary pulsar with a spin period of $\sim 4.8$ s
and an orbital period of $\sim 2.08$ d, discovered with $\uhuru$ 
\cite{Giacconi71}. This X-ray binary is known to have strong long term 
flux variations with transient quasi-periodicity \cite{Priedhorsky83}.
The long term lightcurve of this source obtained with the All Sky Monitor
(ASM) of the {{\it Rossi X-ray Timing Explorer} (\xte)} showed very strong 
aperiodic flux variations along with two different accretion modes
\cite{Paul05}. In 12 years of monitoring data obtained with the {\xte}-ASM
in three different energy bands, the hardness ratio between the 5-12 keV and
3-5 keV bands was found have a larger value during December 2000 to
April 2004, compared to the same outside this period.
 Long term observations with Burst and Transient 
Source Experiment (BATSE) of the {\it Compton Gamma Ray Observatory (CGRO)}
found that Cen X-3 has alternate spin-up and 
spin-down intervals which last from about 10 to 100 days \cite{Finger94}.  
However, {\ginga} observations revealed that there is no correlation between
the observed X-ray flux and pulse period derivative of Cen X-3
\cite{Tsunemi96} and it was suggested that the observed X-ray flux 
of Cen X-3 does not represent its mass accretion rate.

The
periodic long term variations seen in X-ray binaries are explained
by variable obscuration of the central X-ray source by an accretion disk
precessing under tidal force.
The tidal precession can be in the form of forced precession of 
accretion disks in the gravitational field of the companion star \cite{Katz73} 
or slaved precession of accretion disc due to the rotation axis of the star 
being inclined \cite{Roberts74}.
Tilted and twisted disks due to coronal winds \cite{Schandl94,Schandl96}
and radiation pressure induced warped precessing disks
\cite{Iping90,Wijers99,Ogilvie01,Maloney96} can have variable or chaotic
precession periods.  Iping and Petterson (1990) simulated the temporal
evolution of radiatively warped accretion disk and found that it can give
rise to aperiodic variability in the X-ray lightcurves.
They suggested that the long term variability of Cen X-3 could be caused 
by such a radiatively warped accretion disk.
Other more detailed disk warping models have also been developed 
where warping instabilities are incorporated into the models which give rise 
to unstable and chaotic disk precession with no stable long term period 
\cite{Wijers99,Ogilvie01}.
Accretion torque induced precession of magnetic axis can also be a possible
explanation for the observed flux variations \cite{Truemper86} in some
X-ray binaries. Third body mechanism is also known to cause periodic X-ray
flux variation by modulating the mass accretion rate \cite{Zdziarski07b}.

In the present work we investigate the orbital modulation of the 1.5-12 keV
X-ray flux of Cen X-3 when the source is in high, intermediate and
low states to understand if the aperiodic variations are occuring due to 
a variable mass accretion rate or due to the variable obscurtion of the 
central X-ray source by an accretion disk.
The eclipse structure is useful to know the size of the observed X-ray
emission region in different flux states.
We have also measured the pulsed and total X-ray emission of the source
in its different flux states using many observations by
the Proportional Counter Array (PCA) of the $\xte$.
Evolution of the pulsed fraction with observed X-ray flux is useful to
know the relative importance of scattered X-ray emission in different
flux states. 

\section{Observations and Analysis }

The ASM on board $\xte$ has three detectors which scan the sky in a 
series of 90 s dwells in 3 energy bands, namely 1.5-3, 3-5 and 5-12 keV \cite{Levine96} 
. We have used the 1.5-12 keV lightcurves of four 
sources Cen X-3, Her X-1, SMC X-1 and LMC X-4, covering about 4100 days 
from January 1996 to study and compare the orbital modulation of these sources 
in three different flux states. The lightcurves of the 4 sources binned 
with their respective orbital periods after excluding the eclipse data 
are shown in Figure 1, for 500 days. The distinctly aperiodic flux 
variation of Cen X-3 is in sharp contrast with the periodic modulation in
Her X-1 (with one main-on and one short-on state), LMC X-4 (with smaller 
signal to noise ratio) and the quasi-periodic modulation in SMC X-1 (with some 
scatter within each high state).

For each of the three sources Cen X-3, Her X-1 and SMC X-1, we have made
3 separate lightcurves, one each for the high, intermediate and low flux
states. To do this, we first calculated the 
average count rate per orbit after excluding the eclipse data (see Paul et. 
al. 2005 for more details). Depending on the average count rate in a binary 
orbit, the data points available during that orbit were collected in one of 
the three high, intermediate or low state lightcurves (see Table 1 for the 
interval of count rates defining different flux states the three sources).
These 3 lightcurves of each source were then folded with the respective
orbital period of the source to get the orbital modulation lightcurves.

We have adopted a slightly different analysis for LMC X-4 since its signal
to noise ratio is small. We first folded the ASM lightcurve with the
superorbital period and determined the ephemeris for superorbital modulation
as follows.
\begin{equation}
\mathrm{T_{min}[MJD]} = 50092.8 + (30.31\pm 0.01) N,
\label{eq:so_ephemeris}
\end{equation}
where $N$ is an integer. The high, intermediate and low states were determined
based on the suerorbital phase being 0.35 to 0.55, (0.20 to 0.35, 0.55 to 0.80),
and $-$0.20 to 0.20, respectively. Data points from the full ASM lightcurve
belonging to the three different states were then used to make three
different lightcurves and folded with the orbital period of LMC X-4 to get
the orbital modulation lightcurves. 

In Figure 2 we have shown the orbital moduation of the four sources in
different flux states.
For each soure, the orbital modulation lightcurves of the high and
intermediate flux states are shown in the top panel, high state points are
indicated by circles. The low state modulations are shown in the bottom panels.
All the orbital modulation curves are normalised by dividing the original 
lightcurves by the average count rate calculated over an orbital phase of 0.2 
near the peak flux of 
the respective orbital modulation curve. The low state ASM lightcurve for LMC
X-4 does not show any orbtial modulation.
In Table 1 we have also given the orbital periods and eclipse durations 
of these 4 sources determined from the {\xte}-ASM lightcurves. 
In Figure 3, we have shown the flux dependent orbital modulation
lightcurve of Cen X-3 separately for the hard spectral state, that started 
in December 2000 and ended in April 2004 (Paul et al. 2005).

We have also investigated the pulsation characteristics of Cen X-3 at different
flux levels in the 2-60 keV band. Cen X-3 was observed by {\xte}-PCA
many times during the years 1996, 1997 and 1998. All these 
observations have been obtained when Cen X-3 was in the soft spectral state. 
We have chosen 18 PCA observations depending on the orbit averaged ASM 
count rates at that time such that a wide range of X-ray flux is covered. 
The 2-60 keV band lightcurves were obtained from the Standard-1 mode data of
the PCA. Background lightcurves were generated using the background models 
provided for {\xte}-PCA by HEASARC and were subtracted from Standard-1
lightcurves to get the source lightcurves. The source lightcurves were first
barycenter corrected and then searched for the spin period of the neutron star.
We did not detect any pulsations when the orbit averaged ASM count rate of
Cen X-3 was less than 0.8 count/sec (equivalent to about 50 count/sec per 
proportional counter unit)
with a 90 per cent upper limit of 0.8 \% on the pulsed fraction. The pulse
profiles were then generated by folding the barycenter corrected lightcurves
by the respective spin period found in that lightcurve. To avoid 
smearing of the pulse profiles due to the orbital motion of the pulsar, the 
pulse profiles were generated from short data segments of duration of a 
hundred pulses. In Figure 4 we have shown six pulse profiles obtained at
different source flux levels including a lightcurve folded in the
low flux state with a period of 4.81 s when no pulses were detected.
Figure 5 is a plot of the maximum-minimum count rate per Proportional Counter
Unit (PCU) against the maximum count rate per PCU for the 18 pulse profiles.
The points marked with the circles are for the
pulse profiles shown in Figure 4. Epochs of the six PCA observations, pulse
profiles from which are shown in Figure 4, are marked in the top panel
of Figure 1.
A two component function was fitted to the points in Figure 5.

\begin{equation}
F_{\rm max}-F_{\rm min}\simeq \cases{ 0, &$F < F_0$;\cr
f (F_{\rm max} - F_0), &$F\geq F_0$;\cr}
\end{equation} 

The pulse fraction $f$ of the pulsating component was determined
to be 90\%, while the unpulsed component grows upto a count rate of
$F_0$ = 175.5 per PCU.

As seen in Figure 4, apart from the changing ratio of the unpulsed
component of the flux to the total flux, the pulse shape is also varying
from one observation to another. To see whether the pulse shape changes are
related to the flux, we selected three different observations with very
similar eclipse subtracted orbit average ASM count rates of $17.12 \pm 1.35$,
$17.52 \pm 1.21$ and $17.73 \pm 1.47$. Within these three observations, the
pulse shape changed from a double peak profile to a broad single peak profile
but the pulsed fraction remained the same. We therefore conclude that while
the pulsed flux of Cen X-3 is related to the total flux, the pulse shape
is independent of the X-ray flux.

\section{Interpretation and Discussion}

We have found that in the four sources Cen X-3, Her X-1, SMC X-1 and
LMC X-4, the binary orbital modulation of the X-ray flux shows remarkable
dependence on the X-ray flux state (Figure 2).
X-ray eclipses are found to be sharp in the high flux state which
becomes more gradual in the intermediate state.
In the low state, instead of sharp eclipse ingress and egress, there is
a smooth flux variation with orbital phase. Though the orbital modulation
of LMC X-4 is not detectable in the low state of LMC X-4 with ASM data,
we note that a weak and smooth orbital modulation in the low state of LMC X-4
was clearly detected earlier with {\it BeppoSAX} observations
(Naik \& Paul, 2003).  The Her X-1 orbital lightcurves show some pre-eclipse
dips (Figure 2). In the intermediate state of Cen X-3 the eclipse egress
starts earlier in phase by ~0.03 compared to the high state as is shown
with a vertical dashed line.

Using a subset of the {\xte}-ASM lightcurve for Her X-1, Scott \&
Leahy (1999) also found that the orbital modulation is shallower in the short-on
state compared to the same in the main-on state.
In the {\it BeppoSAX} observations of LMC X-4, the absence of clear eclipse
transitions in low state was interpreted as due to obscuration
of the central X-ray source by the precessing accretion disk. The weak
orbital modulation of the lightcurve in the low state is due to an extended
X-ray scttering region, emission from which dominates the detectable X-rays
in the low state.  

The flux dependent orbital modulations of these four sources indicate
that at lower flux, an increasing fraction of the observed X-rays are
from a larger region, comparable to the size of the companion star.
The larger emission region may have different visibility at different orbital
phases, leading to the smooth orbital modulation in the low state.
Reprocessing of X-rays emitted from the compact star by scattering from stellar 
wind of the companion star is one likely scenario.
However, for Her X-1, which has a low mass companion star, the scattering
medium is more likely to be part of the accretion disk and disk outflows than
the stellar wind.
Zdziarski et al (2007a) have perfomed a similar
analysis of {\xte }-PCA lightcurve of the X-ray binary 4U 1820--303 and
found a significant dependence of the profile of the orbital modulation on
the average count rate. However, in 4U 1820--303, the superorbital modulation
is associated with an accretion rate modulation, probably due to third body
interaction.
Poutanen, Zdziarski \& Ibragimov (2008) discovered a superorbital phase
dependence of the soft X-ray orbital modulation in Cyg X-1 in its hard
spectral state, that is related to the size of a bulge in the outer
accretion disk.

The ratio of the X-ray flux when the source is in eclipse and when it is
out-of-eclipse is a measure of the relative scattering efficiency.
Only for Cen X-3 there is good enough statistics to compare the ratios.
We find that the out-of-eclipse count rates differ by large factor ($22.02 \pm 
0.07$, $7.56 \pm 0.02$ and $1.23 \pm 0.02$ for high, intermediate and low 
states respectively) while the in-eclipse count rates of the three states are 
comparable ($0.69 \pm 0.07$ for high, $0.48 \pm 0.03$ for intermediate and 
$0.27 \pm 0.03$ for low state). The eclipse count rate varies by a factor of
only about $\sim$2.5 while the out of eclipse count rate varies by a factor
of $\sim$18.
The ratio of X-ray flux of Cen X-3 during eclipse and out-of-eclipse is 
larger in the low state by a factor of $7.0 \pm 1.3$ compared to the same in
the high state.
This behaviour is similar to SMC X-1, in which the eclipse count rate was
found to be comparable in high and low states whereas the out-of-eclipse count
rate varied by more than a factor of 20 \cite{Wojdowski98}.

The measurement of pulsed X-ray flux as a function of the peak X-ray
flux of Cen X-3 as presented in the Figures 4 and 5, is also consistent
with a scenario in which the measured X-ray flux has two components. One
component is highly variable with a pulsed fraction of about 90\%
and a second component that is unpulsed. Similar result was obtained for 
SMC X-1 over a wide range of its measured X-ray flux \cite{Kaur07}.
In the low state, the unpulsed component becomes dominant
leading to non-detection of pulses. We also note that in all the four sources
mentioned here, the X-ray pulsations (pulse period of $\sim$4.8 s in Cen X-3,
1.24 s in Her X-1, 0.7 s in SMC X-1 and 13.5 s in LMC X-4) have never been
detected during the low state of the superorbital period, indicating that
most of the radiation observed in low state is probably reprocessed emission
from a large region \cite{Wojdowski98,Naik03}. The non-detection of 
pulsations is not due to faintness of the sources. Even in the low state, 
Cen X-3, Her X-1 and SMC X-1 are bright enough for detection of
a few percent pulse modulation with the {\xte}-PCA. 

We point out the possibility that the intrinsic X-ray luminosity of 
the central X-ray source may remain unchanged for a long time.
It can be seen in Figure 1 that except 
during the anomalous low state events of Her X-1, the peak luminosity of 
superorbital modulation of Her X-1, SMC X-1 and LMC X-4 does not change very 
much and the peak of the 5-12 keV X-ray luminosity of Cen X-3 also has a 
ceiling (see Figure 1 of Paul et al. 2005).
Though the results presented here indicate that the different flux 
states of Cen X-3 are largely due to varying degree of obscuration, as is
the case with Her X-1, SMC X-1 and LMC X-4, we cannot completely rule out some 
contribution to the variability from a varying mass accretion rate, especially
the variations associated with spectral change. But as shown in Figure 3, 
even during the hard spectral state of Cen X-3 from December 2000 to April 2004,
the orbital modulation lightcurve indicate an extended X-ray source at
low flux level.
Detailed observations over a wide spectral range with future missions like 
{\it ASTROSAT} can throw more light on 
these aperiodic variations and spectral mode changes seen in Cen X-3 
lightcurves. 

\section{Conclusions}

\begin{enumerate}

\item The binary orbital modulation of X-ray from Cen X-3 is similar to that 
seen in the other three accreting X-ray pulsars. From sharp eclipse in high 
state, it turns to a gradual modulation in the low state. Cen X-3 eclipse 
egress starts earlier in the intermediate state compared to the high state. 
These observations indicate a larger emission region in the low state of Cen 
X-3.
The ratio of X-ray flux of Cen X-3 during eclipse and out-of-eclipse is 
larger in the low state by a factor of $7.0 \pm 1.3$ compared to the same in
the high state.

\item A measurement of the pulsed X-ray flux in different flux states
of Cen X-3 is consistent with the X-ray flux having two components, one with 
a large pulsed fraction and a second unpulsed component that dominates in the 
low state.

\item
We propose that the 
long term intensity variations in Cen X-3 are mostly due to aperiodic 
obscuration of 
the compact source by the accretion disk. The unpulsed X-ray emission from an
extended region appears to be due to scattering of the X-rays from the 
central source by the stellar wind.
\end{enumerate}

\section{Acknowledgements}
We thank the referee A. Zdziarski for very useful suggestions that helped us
very much to improve this paper.
This research has made use of data obtained from the High Energy Astrophysics 
Science Archive Research Center (HEASARC), provided by NASA's Goddard Space 
Flight Centre.

{}

\clearpage
{
\begin{table}
\caption{The source parameters}
\begin{tabular}[b]{|r|r|r|r|r|r|}
\hline
Source&Orbital&Eclipse&\multicolumn{3}{c|}{Orbit averaged}\\
&Period\footnotemark[1]&Duration&\multicolumn{3}{c|}{Count rate for}\\
&(days)&(days)&\\
\hline
      & & &High&Inter-&low\\
      & & &    &mediate&\\   
\hline
Cen X-3&$2.08706 \pm 0.00009$&0.52&$\geq 18.0$&$18.0-2.0$&$\leq2.0$\\
Her X-1&$1.70015 \pm 0.00009$&0.22&$\geq2.5$&$2.5-1.0$&$\leq1.0$\\
SMC X-1&$3.8921 \pm 0.0004$&0.62&$\geq1.3$&$1.3-0.7$&$\leq0.7$\\
LMC X-4\footnotemark[2]&$1.40840\pm 0.00006$&$0.23$&-&-&-\\
\hline
\end{tabular}
\footnote{1}{Orbital periods are measured from {\xte}-ASM lightcurves.}
\footnote{2}{The high, intermediate, and low states of LMC X-4 were
determined using the phases of its stable superorbital period. See \S2
for details.}
\end{table}
}

\clearpage

\begin{figure}
\vskip 2cm
\begin{minipage}{165mm}
\centering
\includegraphics[width=4in,angle=-90]{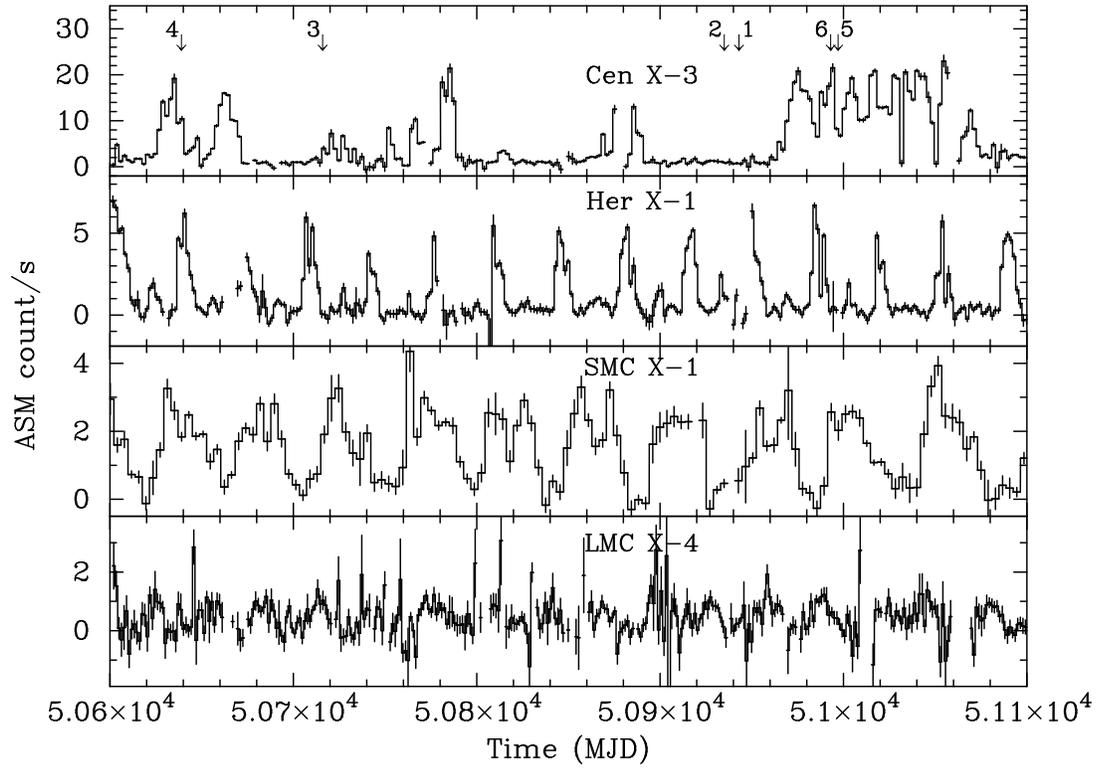}
\caption{The ASM lightcurve of Cen X-3, Her X-1, SMC X-1 and LMC X-4 are shown 
here for 500 days binned with the orbital period of the respective sources. 
The Cen X-3 lightcurve clearly shows aperiodic superorbital variations, 
Her X-1 and LMC X-4 lightcurves show periodic flux variations whereas 
SMC X-1 lightcurve shows quasi periodic flux variations. The numbered arrows
in the Cen X-3 lightcurve represent the times during which the respective
pulse profiles of Figure 4 are taken.}
\end{minipage}
\end{figure}

\clearpage

\begin{figure}
\begin{minipage}{165mm}
\centering
\includegraphics[height=3.2in,width=2.5in,angle=-90]{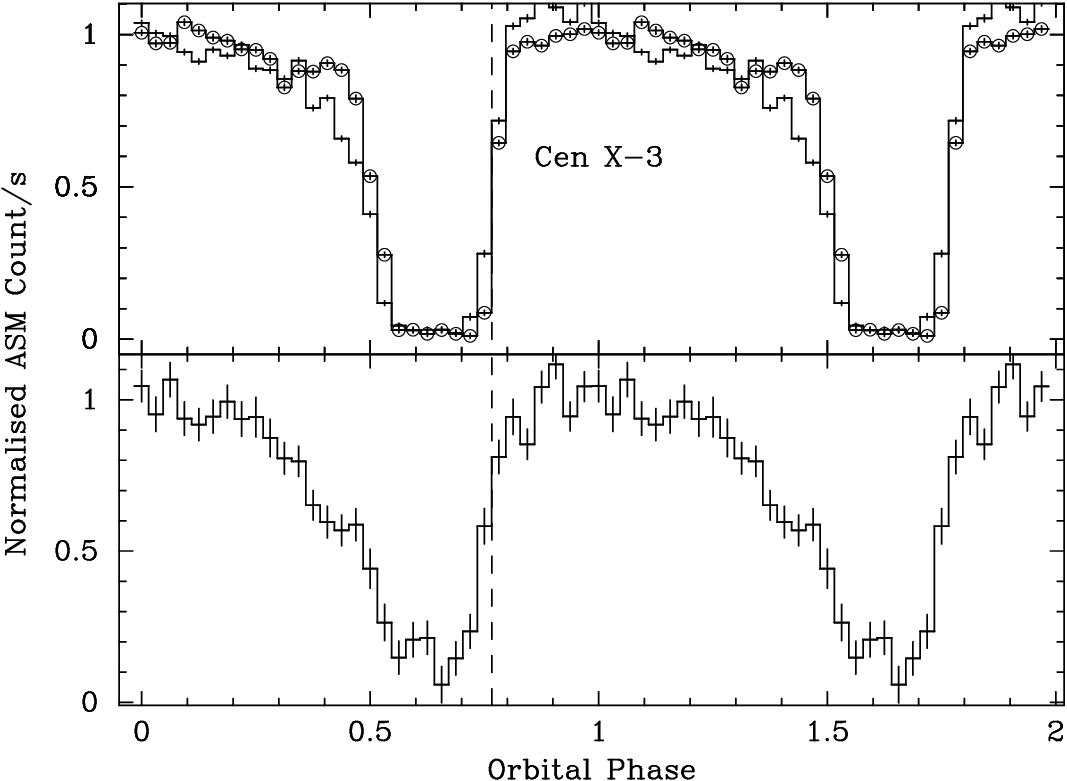}
\includegraphics[height=3.2in,width=2.5in,angle=-90]{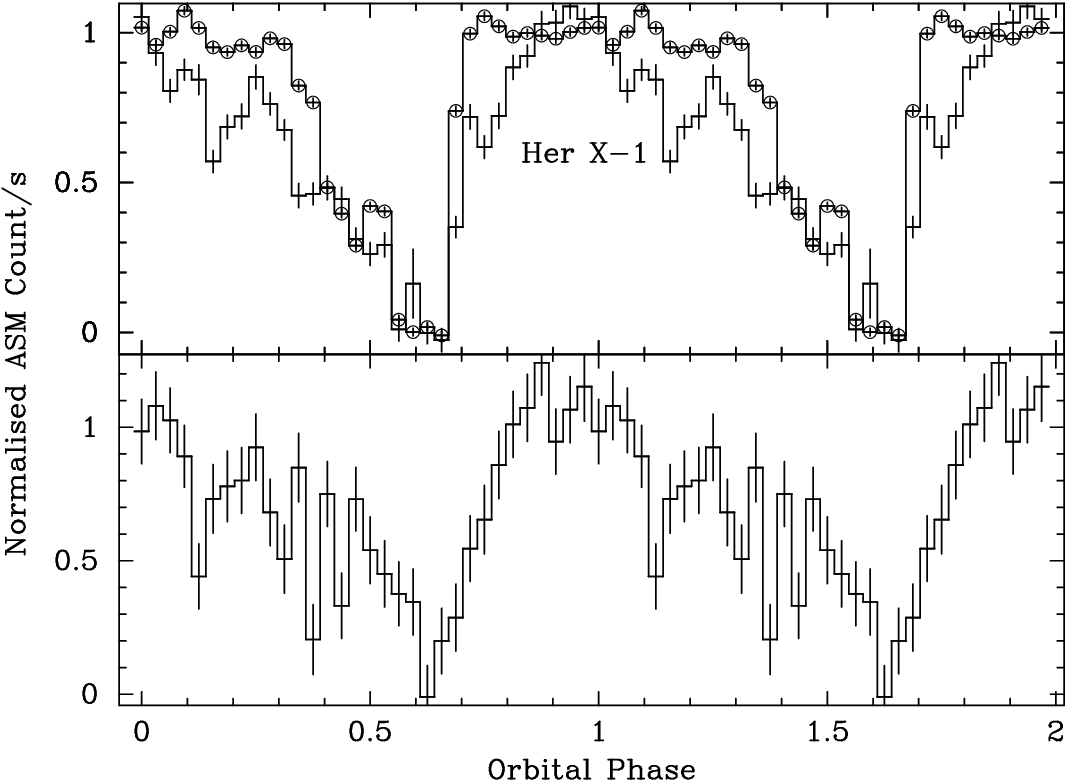}
\includegraphics[height=3.2in,width=2.5in,angle=-90]{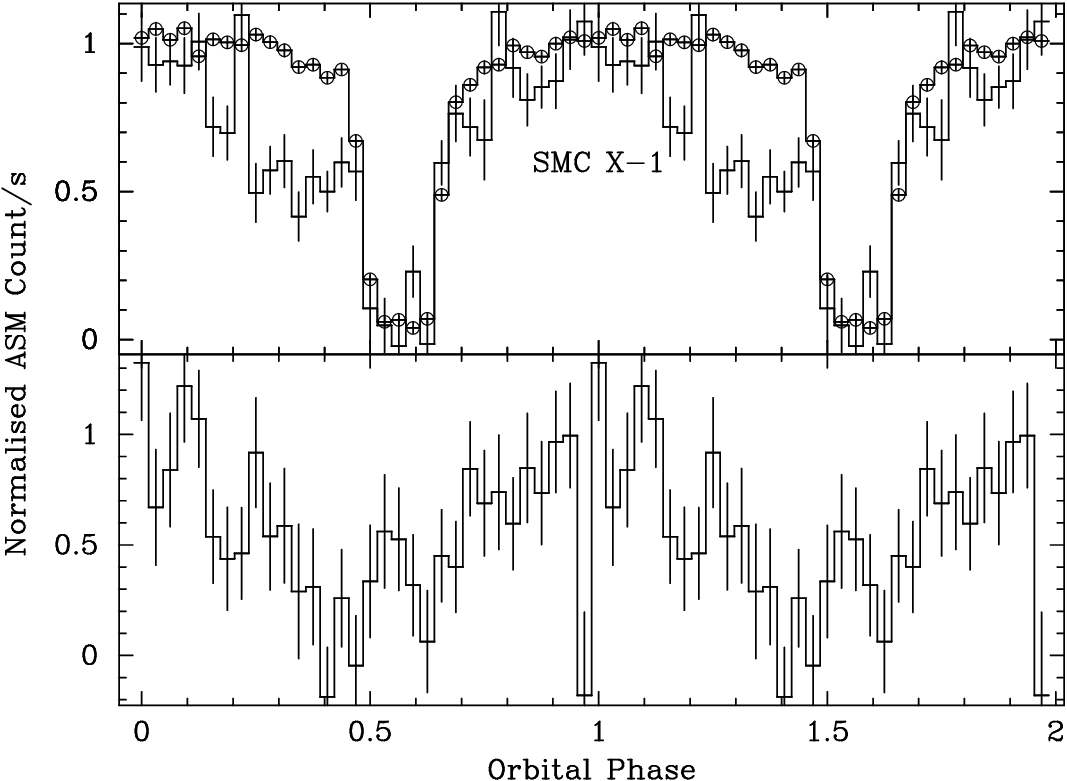}
\includegraphics[height=3.2in,width=2.5in,angle=-90]{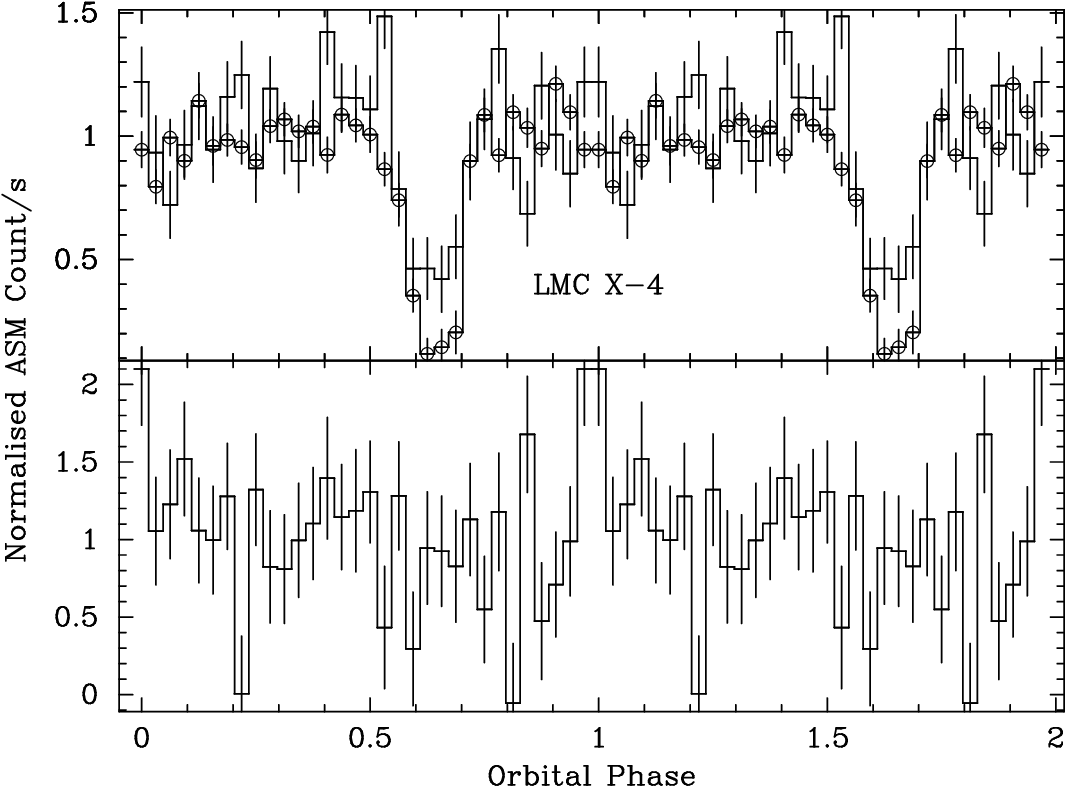}
\caption{The orbital modulation in high, intermediate and low 
states are shown here for four X-ray sources Cen X-3, Her X-1, SMC X-1 and 
LMC X-4. The high and intermediate state plots are shown in the upper pannel,
with the circles denoting the high state points. The plot in the lower pannel 
shows the low state orbital modulation.  
All the plots are normalised by dividing the original curve with the respective
maximum count rate of the curve.   
}
\end{minipage}
\end{figure}

\clearpage

\begin{figure}
\begin{minipage}{165mm}
\vskip 2cm
\centering
\includegraphics[height=5in,angle=-90]{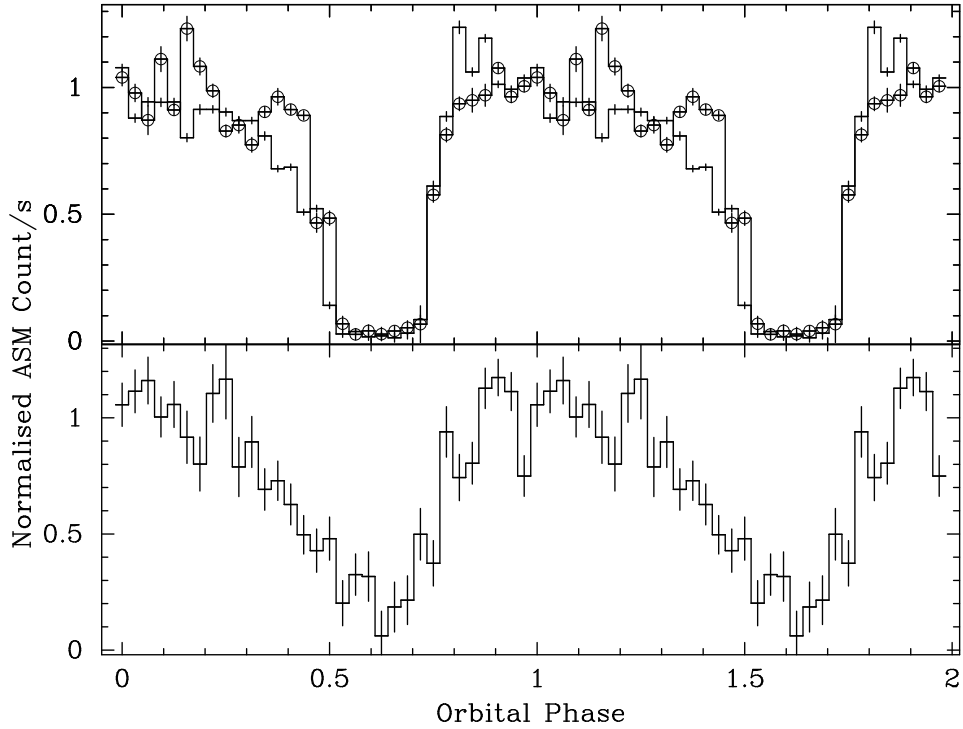}
\caption{Flux dependent orbital modulation of Cen X-3 in hard
spectral state during December 2000 to April 2004. Upper pannel plots show 
high and intermediate state orbital modulation, with high state points being 
marked with circles. Lower pannel plot shows low state orbital modulation.}
\end{minipage}
\end{figure}

\clearpage

\begin{figure}
\begin{minipage}{165mm}
\vskip 2cm
\centering
\includegraphics[height=5in,width=7in,angle=-90]{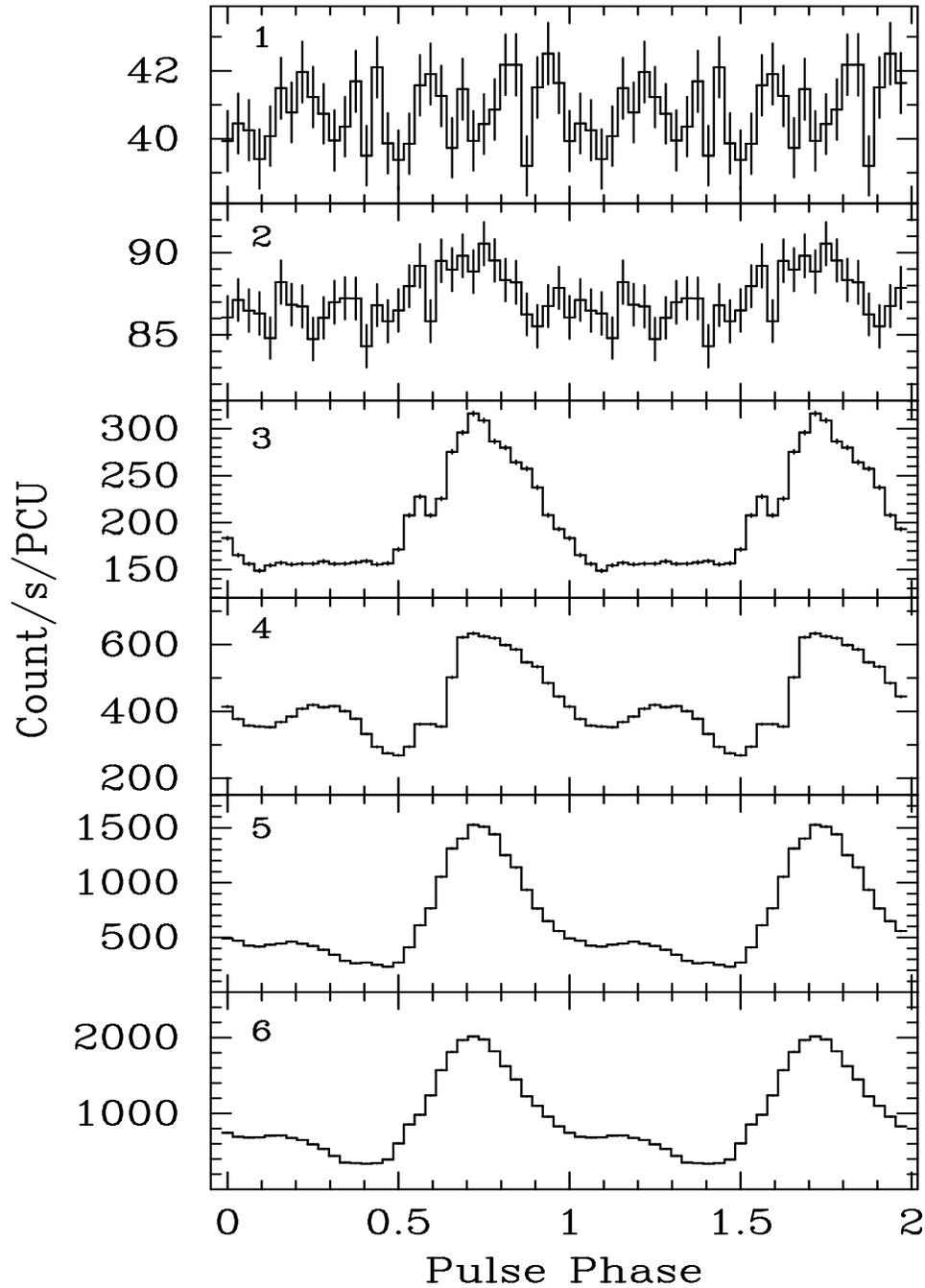}
\caption{Pulse profile of Cen X-3 is shown here in different flux states 
of the source. The count rate is per PCU. Hundred consecutive pulses 
are folded with arbitrary pulse phase and the local spin period to get each of 
the above pulse profiles.}
\end{minipage}
\end{figure}

\clearpage
\begin{figure}
\begin{minipage}{165mm}
\vskip 2cm
\centering
\includegraphics[height=5in, angle=-90]{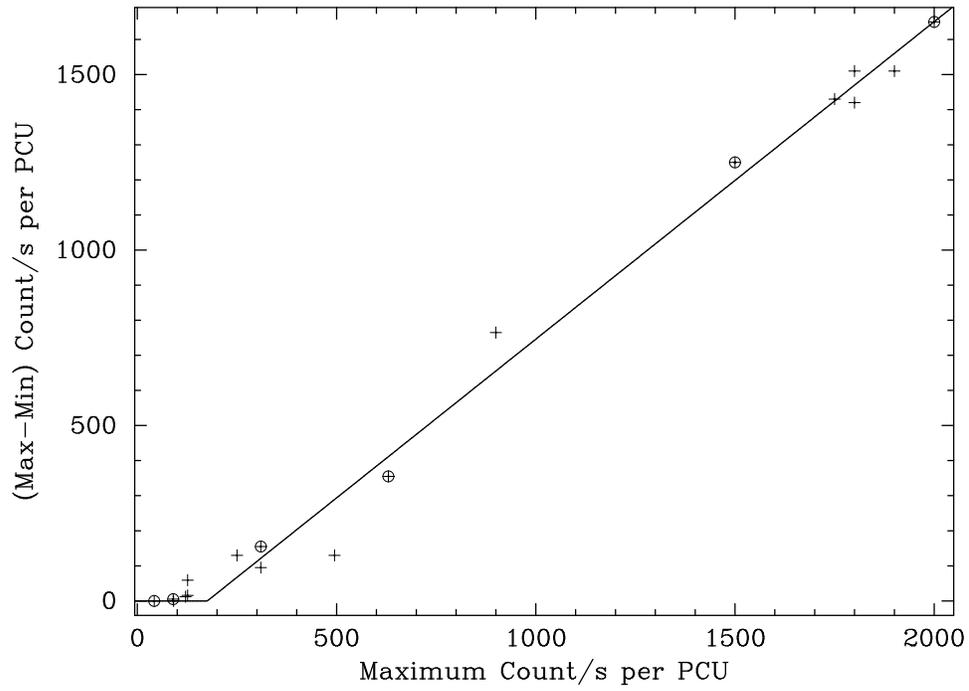}
\caption{Pulsed emission of Cen X-3 is plotted here against the
maximum count rate per detector over a range of X-ray flux of the source.
The points marked with circles correspond to the pulse profiles shown in Figure 4. 
The solid line in the figure is a fit to the function given in the text.}
\end{minipage}
\end{figure}


\begin{thebibliography}{}
\bibitem[{Clarkson et al. }{2003}]{Clarkson03}
Clarkson, W. I., Chales, P. A., Coe, M. J., Laylock, S., Tout, M. D., Wilson, C. A., 2003, MNRAS, 339, 447
\bibitem[{Cornelisse, Charles \& Robertson }{2006}]{Cornelisse06}
Cornelisse, R., Charles, P. A., Robertson, C., 2006, MNRAS, 366, 918
\bibitem[{Eikenberry et al. }{2001}]{Eikenberry01}
Eikenberry, S. S., Cameron, P. B., Fierce, B. W., Kull, D. M., Dror, D. H., Houck, J. R., Margon, B., 2001, ApJ, 561, 1027
\bibitem[{Fabrika }{2004}]{Fabrika04}
Fabrika, S., 2004, ASPRv, 12, 1
\bibitem[{Farrell, Sood \& O'Neill }{2006}]{Farrell06}
Farrell, S. A., Sood, R. K., O'Neill, P. M.,  2006, MNRAS, 367, 1457
\bibitem[{Finger, Wilson \& Fishman }{1994}]{Finger94}
Finger, M. H., Wilson, R. B., Fishman, G. J., 1994, Second Compton Symposium,p 304
\bibitem[{Giacconi et al. }{1971}]{Giacconi71}
Giacconi, R., Gursky, H., Kellogg, E., Schreier, E., Tananbaum, H., 1971, ApJ, 167, 67
\bibitem[{Gruber \& Rothschild }{1984}]{Gruber84}
Gruber, D. E., Rothschild, R. E., 1984, ApJ, 283, 546
\bibitem[{in't Zand et al. }{2003}]{Zand03}
in't Zand, J. J. M., Hulleman, F., Markwardt, C. B., Mendez, M., Kuulkers, E., Cornelisse, R., Heise, J., Strohmayer, T. E., Verbunt, F., 2003, A\&A, 406, 233 
\bibitem[{Iping \& Petterson }{1990}]{Iping90}
Iping, R. C., Petterson, J. A., 1990, A\&A, 239, 221
\bibitem[{Katz }{1973}]{Katz73}
Katz, J. I., 1973, Nature Physical Sciences, 246, 87
\bibitem[{Kaur et al. }{2007}]{Kaur07}
Kaur, R., Paul, B., Raichur, H., Sagar, R., 2007, ApJ, 660, 1409
\bibitem[{Kitamoto et al., }]{Kitamoto00}
Kitamoto, S., Egoshi, W., Miyamoto, S., Tsunemi, H., Ling, J. C., Wheaton, W. A., Paul, B., 2000, 531, 546
\bibitem[{Lachowicz et al. }{2006}]{Lachowicz06}
Lachowicz, P., Zdziarski, A. A., Schwarzenberg-Czerny, A., Pooley, G. G.,
Kitamoto, S., 2006, MNRAS, 368, 1025
\bibitem[{Levine et al. }{1996}]{Levine96}
Levine, A. M., Bradt, H., Cui, W., Jernigan, J. G., Morgan, E. H., Remillard, R., Shirey, R. E., Smith, D. A., 1996, ApJ, 469, 33
\bibitem[{Maloney, Begelman \& Pringle }{1996}]{Maloney96}
Maloney, P. R., Begelman, M. C., Pringle, J. E., 1996, ApJ, 472, 582
\bibitem[{Masetti }{2002}]{Masetti02}
Masetti, N. 2002, A\&A, 381, L45
\bibitem[{Naik \& Paul }{2003}]{Naik03}
Naik, S., Paul, B., 2003, A\&A, 401, 265
\bibitem[{Ogilvie \& Dubus }{2001}]{Ogilvie01}
Ogilvie, G. I., Dubus, G., 2001, MNRAS, 320, 485
\bibitem[{Paul \& Kitamoto }{2002}]{Paul02}
Paul, B., Kitamoto, S., 2002, JAA, 23, 33
\bibitem[{Paul, Kitamoto \& Makino }{2000}]{Paul00}
Paul, B., Kitamoto, S., Makino, F., 2000, ApJ, 528, 410 
\bibitem[{Paul, Raichur \& Mukherjee }{2005}]{Paul05}
Paul, B., Raichur, H., Mukherjee, U., 2005, A\&A, 442, L15
\bibitem[]{}
Poutanen, J, Zdziarski, A. A., Ibragimov, A. 2008, arXiv0802.1391, Submitted to MNRAS
\bibitem[{Priedhorsky \& Terrell }{1983}]{Priedhorsky83}
Priedhorsky, W. C.; Terrell, J., 1983, ApJ, 273, 709
\bibitem[{Roberts }{1974}]{Roberts74}
Roberts, W. J., 1974, ApJ, 187, 575
\bibitem[{Schandl }{1996}]{Schandl96}
Schandl, S., 1996, A\&A, 307, 95
\bibitem[{Schandl \& Meyer }{1994}]{Schandl94}
Schandl, S., Meyer, F., 1994, A\&A, 289, 149
\bibitem[{Scott \& Leahy }{1999}]{Scott99}
Scott, D. M., Leahy, D. A., 1999, ApJ, 510, 974 
\bibitem[{Smale \& Lochner }{1992}]{Smale92}
Smale, A. P., Lochner, J. C., 1992, ApJ, 395, 582 
\bibitem[{Still \& Boyd }{2004}]{Still04}
Still, M., Boyd, P., 2004, ApJ, 606, L135.
\bibitem[{Tananbaum et al. }{1972}]{Tananbaum72}
Tananbaum, H., Gursky, H., Kellogg, E. M., Levinson, R., Schreier, E.,
Giacconi, R., 1972, ApJ, 174, L143
\bibitem[{Truemper et al. }{1986}]{Truemper86}
Truemper, J., Kahabka, P., Oegelman, H., Pietsch, W., Voges, W., 1986, ApJ, 300, 63
\bibitem[{Tsunemi, Kitamoto \& Tamura }{1996}]{Tsunemi96}
Tsunemi, H., Kitamoto, S., Tamura, K., 1996, ApJ, 456, 316
\bibitem[{Wen et al. }{2006}]{Wen06}
Wen, L., Levine, A. M., Corbet, R. H. D., Bradt, H. V., 2006, ApJS, 163, 372
\bibitem[{Wijers \& Pringle }{1999}]{Wijers99}
Wijers, R. A. M. J., Pringle, J. E., 1999, MNRAS, 308, 207
\bibitem[{Wojdowski et al. }{1998}]{Wojdowski98}
Wojdowski, P., Clark, G. W., Levine, A. M., Woo, J. W., Zhang, S. N., 1998, ApJ, 502, 253
\bibitem[{Zdziarski et al. }{2007a}]{Zdziarski07a}
Zdziarski, A. A., Gierli\'nski, M., Wen, L., Kostrzewa, Z., 2007a, MNRAS, 377, 1017
\bibitem[{Zdziarski, Wen \& Gierli\'nski }{2007b}]{Zdziarski07b}
Zdziarski, A. A., Wen, L., Gierli\'nski, M., 2007b, MNRAS, 377, 1006
\end{thebibliography}
\end{document}